# 1D Quantum transport in the even-chain spin-ladder compound $Sr_{2.5}Ca_{11.5}Cu_{24}O_{41}$ and $YBa_2Cu_4O_8$

V.V. Moshchalkov, L. Trappeniers and J. Vanacken

*Laboratorium voor Vaste-Stoffysica en Magnetisme,
Katholieke Universiteit Leuven,
Celestijnenlaan 200 D, B-3001 Leuven, Belgium*

The temperature dependence of the resistivity $r(T)$ of the novel $Sr_{2.5}Ca_{11.5}Cu_{24}O_{41}$ ladder compound under hydrostatic pressure of up to 8 GPa has been explained by assuming that the relevant length scale for electrical transport is given by the magnetic correlation length related to the opening of a spin-gap in a 1D even-chain spin-ladder (1D-SL). The pressure dependence of the gap was extracted by applying this model to $r(T)$ data obtained at different pressures. The $r(T)$ dependence of the underdoped cuprate $YBa_2Cu_4O_8$ demonstrates a remarkable scaling with the $r(T)$ of the 1D-SL compound $Sr_{2.5}Ca_{11.5}Cu_{24}O_{41}$. This scaling implies that underdoped cuprates at $T_c < T < T^*$ are in the 1D regime and their pseudo-gap below $T^*$ is the spin-gap in the even-chain 1D-SL formed at $T < T^*$ in these materials.

The temperature dependent resistivity $r(T)$ of high-$T_c$ cuprates has recently been interpreted in the framework of the two-dimensional (2D) Heisenberg model [1]. The approach proposed in Ref. 1 is based on the three basic assumptions: (i) the dominant scattering mechanism in high-$T_c$'s in the whole temperature range is of *magnetic origin*; (ii) the specific temperature dependence of the resistivity $r(T)$ can be described by the *inverse quantum conductivity* $s^{-1}$ with the inelastic length $L_f$ being fully controlled, via a strong interaction of holes with $Cu^{2+}$ spins, by the *magnetic correlation length* $x_m$, and, finally, (iii) the proper *2D expressions* should be used for calculating the quantum conductivity with $L_f \sim x_m$.

With a few straightforward modifications, these ideas can also be applied to the 1D quantum spin-ladder (SL) systems [2]. In this case the quantum resistance of a single 1D wire is a linear function of the inelastic length $L_f$ [3] and $r(T)$ can be represented as

$$r_{1D}^{-1}(T) = s_{1D}(T)\big|_{L_f = x_m} = \frac{1}{b^2}\frac{e^2}{\hbar} x_{m1D}(T) \qquad (1)$$

Here $b$ is the diameter of a single 1D wire and $L_f \sim x_{m1D}$ is given by the spin-correlation length $x_{m1D}$ in the 1D regime. For a quantum 1D transport, the concentration of the charge carriers does not enter directly into the prefactor in Eq. 1. Instead, indirectly, the Fermi wavelength $l_F$ (and therefore also the concentration) defines the number of channels available for the quantum transport. The Landauer-Büttiker formalism [4] describes quantum conductivity just by calculating the number of the contributing channels. As was clearly demonstrated by recent experiments with ultrathin metallic wires [5], quantized conductance indeed follows the predictions of the 1D quantum metallic transport. Therefore, for a one-channel quantum transport Eq. 1 can be used without any explicit dependence on the charge carrier concentration. For a collection of $N_{(1D)}$ 1D-wires, connected as parallel independent conducting paths, $\sigma_{1D}$ (Eq. 1) should be multiplied by $N_{(1D)}$. If we assume that the 1D conducting paths are formed in the Cu-O planes in high $T_c$ cuprates as a result of doping, then $r(T)$ becomes concentration dependent via $N_{(1D)} \sim p$, where $p$ is the concentration of holes.

To use Eq. 1 for 1D even-chain Heisenberg SL compounds the spin-correlation length found by Monte Carlo simulations [6] can be taken for $x_{m1D}$:

$$(\Delta x_{m1D})^{-1} = \frac{2}{p} + A\left(\frac{T}{\Delta}\right)\exp\left(\frac{-\Delta}{T}\right) \tag{2}$$

where $A \approx 1.7$ and $D$ is the spin-gap. The combination of Eq. 1 and Eq. 2 leads to

$$r(T) = r_0 + CT\exp\left(-\frac{\Delta}{T}\right) = \frac{\hbar b^2}{e^2 a}\left\{\frac{2\Delta}{pJ_{//}} + A\frac{T}{J_{//}}\exp\left(-\frac{\Delta}{T}\right)\right\} \tag{3}$$

where $J_{//}$ is the intra-chain coupling and $a$ is the spacing between the 1D wires. This $r(T)$ dependence of the 1D system with $L_f \sim x_{m1D}$ can be directly used to analyze available transport data reported recently for metallic SL compounds, which are clearly in the 1D regime due to their specific crystalline structure. If one considers the possibility of stripe formation in the Cu-O planes, then doping of this presumably 2D system might occur in a highly inhomogeneous manner, thus reducing the symmetry from 2D to 1D and producing conducting 1D paths separated by the distance $a \sim 1/p$. From that point of view, one should expect the scaling of the product $\rho(T) \cdot p$, if doping leads to the formation of an increasing number ($\sim p$) of *identical* stripes. The latter is not evident a priori, since, in principle, such parameters as $\Delta$, $J_{//}$ and $b$ (Eq. 3) might be also concentration dependent. The plot of the product $r(T) \times p$ has been already used in Ref. 1 ($x = p$), where a nice scaling has been observed for the $r(T) \times p$ curves of the underdoped cuprates.

To verify the validity of the proposed 1D SL model $\sigma_{1D} \sim \xi_{m1D}$, we use it first for the description of the resistivity data obtained on single crystals of the novel even-chain SL compound $Sr_{2.5}Ca_{11.5}Cu_{24}O_{41}$ [7]. This compound definitely contains a two-leg ($n_c = 2$) $Cu_2O_3$ ladder and therefore its resistivity along the ladder direction should indeed obey Eq. 1-3 with $x_{m1D}$ given by the recent Monte Carlo calculations [6], through the admixture of the linear temperature dependence of $x_{m1D}^{-1}$ with the exponential term containing the spin gap $\Delta$ and the constant A (see Eq. 2). The results of the $r(T)$ fit with Eq. 3 are shown in Fig. 1. This fit demonstrates a remarkable quality over the whole temperature range $T \approx 25\text{-}300$ K, except for the lowest temperatures where the onset of the localisation effects, not considered here, is clearly visible in the experiment [7]. Moreover, the used fitting parameters $r_0$, $C$ and $\Delta$ in Eq. 3 all show very reasonable values. The expected residual resistance $r_0 = \dfrac{\hbar b^2}{e^2 a}\dfrac{2\Delta}{pJ_{//}}$ for $b \sim 2a \sim 7.6$ Å, $\Delta \sim 200$ K and $J_{//} \sim 1400$ K (the normal value for the CuO$_2$ planes) is $r_0 \approx 0.5\cdot 10^{-4}$ Ωcm which is in good agreement with $r_0 \approx 0.83\cdot 10^{-4}$ Ωcm found from the fit. The fitted gap $\Delta \approx 216$ K (at 8 GPa) (Fig. 1) is close to $\Delta \approx 320$ K determined for the undoped SL SrCu$_2$O$_3$ from inelastic neutron scattering experiments [8]. In doped systems it is natural to expect a reduction of the spin gap. Therefore the difference between the fitted value (216 K) and the one measured in an undoped system (320 K) seems to be quite fair. Finally the calculated fitting parameter $C = \dfrac{r_0 A p}{2\Delta} = 0.0103$ (in units of $10^{-4}$ Ωcm/K) is to be compared with $C = 0.013$ (from the 8 GPa fit in Fig. 1). Using the fitting procedure for the two pressures 4.5 GPa ($\Delta \approx 219$ K) and 8 GPa ($\Delta \approx 216$ K), we have obtained a weak suppression of the spin-gap under pressure $d\Delta/dp \sim -1$ K/GPa. The use of other published single crystal data [9], results in a fit of a similar quality.

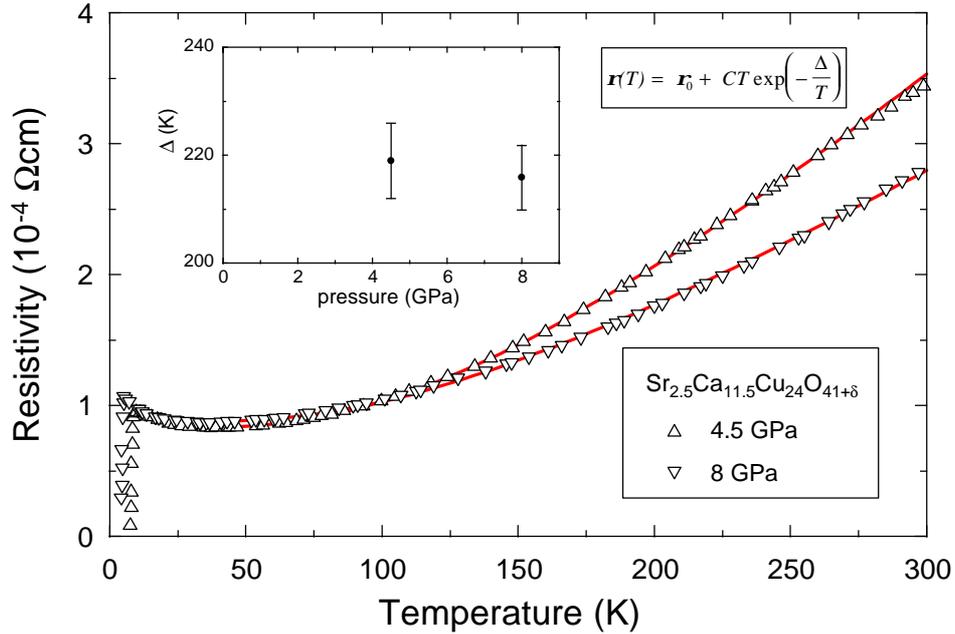

*Fig. 1 Temperature dependence of the resistivity for a Sr$_{2.5}$Ca$_{11.5}$Cu$_{24}$O$_{41}$ even-chain spin-ladder single crystal along the spin chains at 4.5 GPa and 8 GPa (experimental data points after Ref. 7). The solid line represents a fit using Eq. 3 describing transport in 1D SL's.*

Since the spin-gap is the gap for magnetic excitations, the number $N_m$ of the available magnetic scatterers (magnons, etc.) will be determined by the thermal excitation across the gap $\Delta$: $N_m \sim exp(-\Delta/T)$. In this case $L_f$, as the average distance between magnetic scatterers, is also an exponential function of $\Delta/T$. This explains a quite general nature of the exponential factor in Eq. 2 - Eq. 3. It is, however, not clear how one should interpret the pre-exponential $T$ factor in simple terms. At the same time one should keep in mind that this factor is essential since it gives the correct linear temperature dependence $\Delta \rho(T) \propto T(1-\Delta/T)$ at high temperatures $T > \Delta$.

A rapidly growing experimental evidence [10-17] indicates that the 1D scenario might be also relevant for the description of the underdoped high-$T_c$ cuprates where 1D stripes can be eventually formed. Since mobile carriers in this case are expelled from the surrounding Mott insulator phase into the stripes, the latter then provide *the lowest resistance paths. This makes the transport properties very sensitive to the formation of the stripes - both static and dynamic.* We assume that for the 1D stripes the condition $L_f \sim \xi_{m1D}$ is enforced either by the proximity effect between a metallic stripe and the Mott insulator phase [18] (see also Fig. 2, inset) or by the effective regime of the doped even-chain SL's. The latter implies that conducting stripes could be just doped ladders, spontaneously formed in the CuO planes under doping.

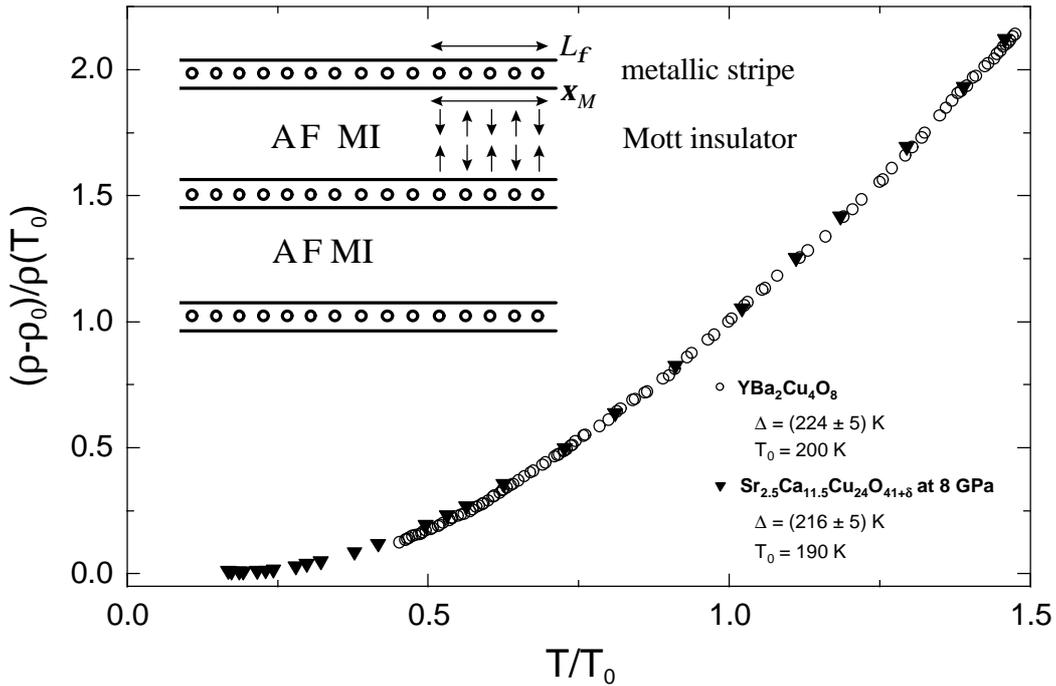

*Fig. 2 Scaled temperature dependence of the resistivity for a YBa$_2$Cu$_4$O$_8$ single crystal (open circles) and a Sr$_{2.5}$Ca$_{11.5}$Cu$_{24}$O$_{41}$ even-chain SL single crystal at 8 GPa (filled down-triangles). The inset shows a schematic view of the formation of 1D metallic stripes surrounded by an antiferromagnetic Mott insulator. $L_F$ denotes the inelastic length and $\xi_m$ the magnetic correlation length.*

To investigate the possibility of using the 1D scenario for describing transport properties of the 2D Cu-O planes, we firstly directly compare the temperature dependencies of the resistivity of a typical underdoped high-$T_c$ material $YBa_2Cu_4O_8$ with that of the even-chain SL compound $Sr_{2.5}Ca_{11.5}Cu_{24}O_{41}$. The crystal structure of the $YBa_2Cu_4O_8$ compound ('124') differs substantially from that of the more common $YBa_2Cu_3O_7$ ('123'), since 124 contains double CuO chains stacked along the c-axis and shifted by *b/2* along the b axis [19]. These chains are believed to act as charge reservoirs, therefore they may have a strong influence on the transport in the $CuO_2$ planes themselves. In the 124 case, the 1D features of this double CuO chain can be expected to induce an intrinsic doping inhomogeneity in the neighbouring $CuO_2$ planes thus enhancing in a natural way the formation of the 1D stripes. A weak coupling of 1D chains to 2D planes might be sufficient to reduce the effective dimensionality by orienting the stripes in the $CuO_2$ planes along the chains. But even in pure 2D planes, without their coupling to the 1D structural elements, the formation of the 1D stripes seems to be possible.

Early experiments on twinned high-$T_c$ samples have created an illusion that all planar Cu sites in the $CuO_2$ planes are equivalent. Recent experiments [20-26] on perfect untwinned single crystals have completely unveiled this illusion. A very large anisotropy in the *ab*-plane of twin-free samples has been reported for resistivity ($r_a/r_b(YBa_2Cu_2O_7)$) = 2.2 (Gagnon et al [20], Friedman et al [21]), $r_a/r_b(YBa_2Cu_4O_8)$ = 3.0 (Bucher and Wachter [22]), thermal conductivity ($k_a/k_b(YBa_2Cu_4O_8)$) = 3-4 (Cohn and Karpinski [23])), for superfluid density [24,25] and optical conductivity [25,26]. In all these experiments, much better metallic properties have been clearly seen along the direction of the chains (*b*-axis). And what is truly remarkable, that this large in-plane anisotropy is strongly suppressed by a small (only 0.4 %) amount of Zn [25], which is known to replace copper, at least for Zn concentrations up to 4 %, only in the $CuO_2$ planes ! The latter suggests that the *ab*-anisotropy can not be explained just by assuming the existance of highly conducting CuO-chains. Instead, this observation highlights the fact that the in-plane anisotropy is caused by certain processes in the $CuO_2$ planes themselves, where the substitution of Cu by Zn takes place. In this situation we may expect that the chains are actually imposing certain directions in the $CuO_2$ planes for the formation of the 1D stripes.

This expectation is supported by our observation of a perfect scaling of the *r(T)* curves for 1D spin ladders and (presumably "2D") $YBa_2Cu_4O_8$. Using a simple scaling parameter $T_0$, we have found a convincing overlapping of the two sets of data: *(r-r₀)/r(T₀)* versus *T/T₀* (with *r₀* being the residual resistance) for $YBa_2Cu_4O_8$ and $Sr_{2.5}Ca_{11.5}Cu_{24}O_{41}$. Note that *r₀* should be subtracted from *r(T)* since *r₀* may contain contributions from several scattering mechanisms depending on the sample quality. The remarkable scaling behaviour (Fig. 2) demonstrates that *resistivity vs. temperature dependencies of underdoped cuprates in the pseudo-gap regime at $T < T^*$ and even-chain SL with the spin-gap **D** are governed by the same underlying mechanism*. Since the 1D quantum conductivity model (Eq. 3) works very well for even-chain SL's (see Fig. 1), then it should be also applicable for the description of the *r(T)* behaviour in the underdoped cuprates. The fitting of the *r(T)* curve for $YBa_2Cu_4O_8$ with Eq. 3 has resulted in a spin-gap $\Delta = (224 \pm 5)\ K$ (Fig. 3). Therefore, the resistivity of underdoped cuprates below $T^*$ simply reflects the temperature dependence of the magnetic correlation length $\chi_{m1D} \propto r^{-1}$ (Eq. 2) in the even-chain SL's (see Fig. 5 in Ref. 6) and the pseudo-gap is the spin-gap formed in the 1D stripes.

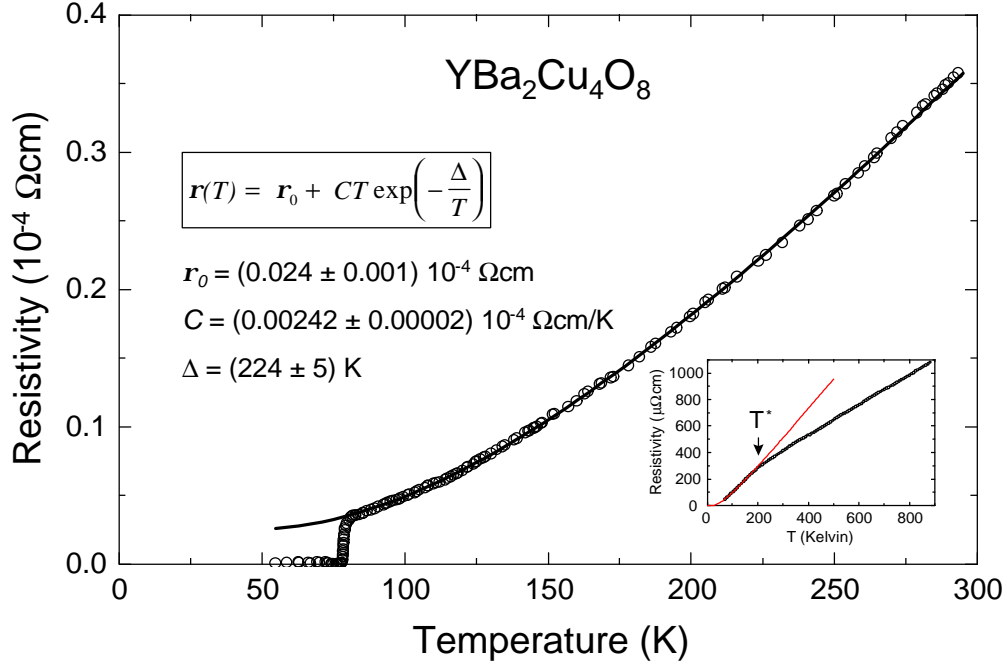

*Fig. 3   Temperature dependence of the resistivity for a YBa$_2$Cu$_4$O$_8$ single crystal (open circles); the solid line represents the fit using Eq. 3. The high-temperature data taken on another crystal (after Ref. 19), shown in the inset, illustrate the crossover at $T^*$ to 2D (linear behaviour).*

It seems to be quite important to support these observations and conclusions by applying similar ideas in the analysis of other physical properties. Since in underdoped cuprates the scaling temperature $T_0$ works equally well for resistivity and Knight-shift data [27,28], the latter can also be used for fitting with the expressions derived from the 1D SL models. Then the temperature dependence of the Knight shift should obey the following expression [29]:

$$K(T) \sim T^{-\frac{1}{2}} \exp\left(-\Delta/T\right) \qquad (4).$$

Fitting the *K(T)* data [27] for YBa$_2$Cu$_4$O$_8$ with Eq. 4 gives an excellent result (Fig. 4) with the spin-gap $\Delta = (222 \pm 20)\,K$ which is very close to the value $\Delta = (224 \pm 5)\,K$ derived from the resistivity data (Fig. 3).

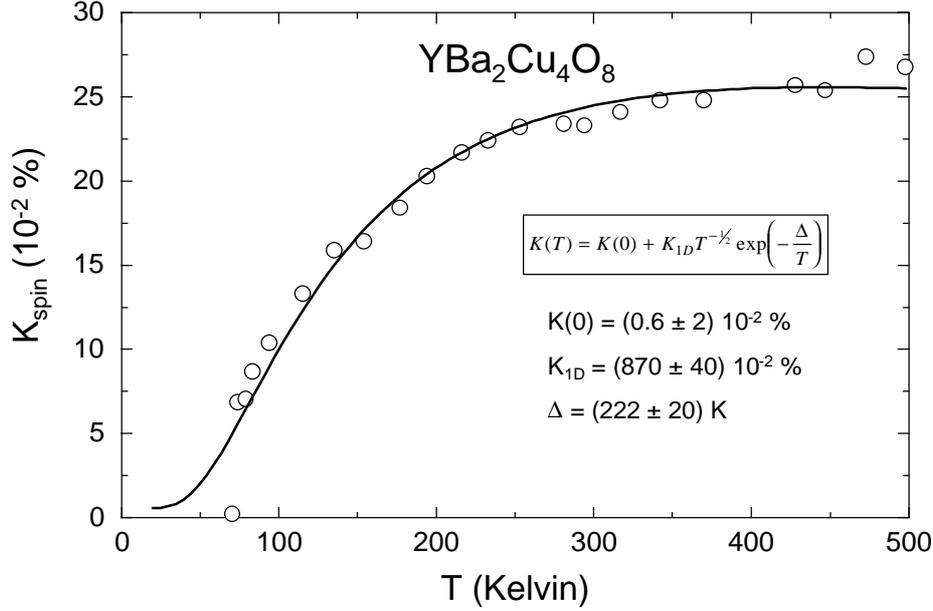

*Fig. 4 Knight shift data (open circles) (after Ref. 19) fitted with Eq. 4 (solid line). A gap of* $\Delta = (222 \pm 20)\ K$ *was obtained.*

The proposed approach to the description of the scaling behaviour of the resistive properties of cuprate high-$T_c$'s and the existence of the pseudo-gap is based on the assumption that DC transport in high-$T_c$'s is caused by the quantum conductivity with the inelastic length $L_f$ fully governed by the magnetic correlation length $x_m$. Depending upon the effective dimensionality - 2D ($T > T^*$) or 1D (stripes at $T < T^*$) (Fig. 5) - conductivity is found to change from $s_{2D} \sim ln(x_{m2D})$ [1] to $s_{1D} \sim x_{m1D}$ [2] (Eq. 1), respectively. The validity of the $s_{1D} \sim x_{m1D}$ approach has been checked directly using resistivity data for real SL systems. The temperature dependence $r(T)$ of single crystals of the novel $Sr_{2.5}Ca_{11.5}Cu_{24}O_{41}$ ladder compound under hydrostatic pressure of up to 8 GPa [7] and the underdoped cuprate $YBa_2Cu_4O_8$ have been explained in the framework of the model assuming the opening of a spin-gap in the 1D even-chain spin-ladder (1D-SL) and a pseudo-gap at $T < T^*$ in $YBa_2Cu_4O_8$. The $r(T)$ dependence of the underdoped cuprate $YBa_2Cu_4O_8$ also demonstrates a remarkable scaling with the $r(T)$ of the 1D-SL compound. This scaling implies that the pseudo-gap below $T^*(p)$ ($p$ being the hole concentration) in underdoped $YBa_2Cu_4O_8$ is the spin-gap in the even-chain SL's formed at $T < T^*(p)$ in these materials (Fig. 5) and strongly reflects the 1D nature of their microscopic electronic structure in this regime. Other possibilities to fulfil the essential constraint of our model ($L_f \sim x_{m1D}$), are proximity like effects [11] between the Mott insulating 1D domain (with $x_{m1D}$ controlled by the SL model) and conducting 1D stripes. In this case, due to the separation of metallic and insulating domains the $x_{m1D}$ behaviour in doped cuprates will be very similar to that in underdoped insulating 1D - SL compounds. Moreover, the dynamic character of the stripe formation will inevitably produce an *averaged* spin dynamics which is seen in neutron scattering experiments on doped cuprates. Contrary to that *no averaging of the transport properties will occur*, since, even for dynamic stripes, the charge will automatically follow the most conducing paths, i.e. stripes, even if they are moving fast. Therefore, the spin-spin correlations seen in neutron scattering experiments [13,17] correspond to the picture averaged over dynamic insulating and conducting stripes, whereas in transport experiments the magnetic correlation length $x_{m1D}$ of a dynamic insulating stripe permanently imposes the constraint $L_f \sim x_{m1D}$, thus providing the persistent 1D character of the charge transport in underdoped cuprates.

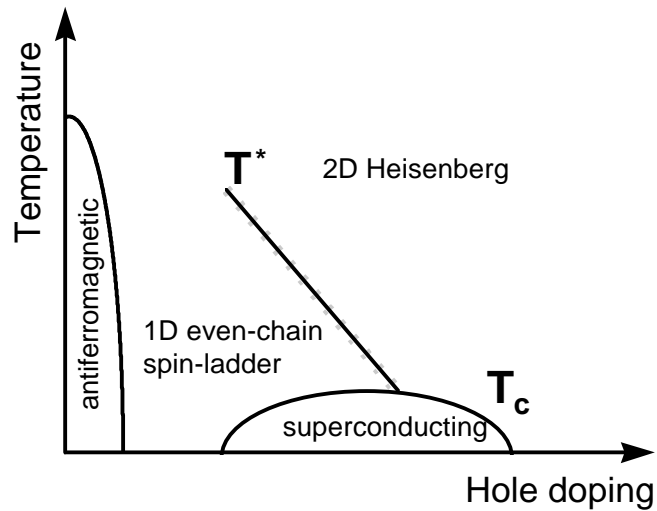

*Fig. 5   Schematic phase diagram of layered high-$T_c$ cuprates.*

The proposed approach considers the pseudo-gap temperature $T^*(p)$ as the crossover between the high temperature 2D Heisenberg and the 1D quantum even-chain SL regime. The latter is established as temperature decreases and charge expulsion from the Mott insulator phases [30] gets stronger and finally leads to the formation of well defined separate metallic and insulating stripes which can be modelled by the SL with an even number of chains. The effective 1D case at $T < T^*(p)$ makes all considerations of the non-Fermi liquid behaviour [11,31,32] and possible charge/spin separation [33,34] important for the physics of underdoped high-$T_c$'s in general. The weakly doped even-chain SL's have the spin-gap, show hole-hole pairing (though rather short-range along the chain) with an "approximate" $d_{x^2-y^2}$ symmetry [14]. These materials, being lightly doped insulators, show also features of a large metal-like Fermi surface. Since the 1D SL phase precedes the superconducting transition, then it is quite reasonable to consider similar mechanisms being responsible for superconductivity in both underdoped cuprates and in SL with an even number of chains [14]. The short-range correlation for the pairs along the chains in even-leg SL's [35] contradicts ARPES and corner-junction SQUID measurements carried out at $T < T_c(p)$. This can be explained by a substantial low-temperature modification of these correlations due to the onset at $T \sim T_c(p)$ of the Josephson-like coupling between the stripes [11], resulting eventually in an effective recovery of the 2D-character of the CuO system at $T < T_c(p)$. The Josephson-like coupling between the 1D SL causes the onset of superconductivity at $T_c(p)$ with the $T_c(p)$ value increasing with the hole doping. Isolated 1D even-chain SL's, most probably, have no chance to develop along the chain a real macroscopic superconducting coherence of bosons, formed along rungs in the SL. From this point of view, recent pulsed field data [36] can be interpreted as an experimental evidence of the insulating ground state of field-decoupled SL's in underdoped cuprates.

The 1D SL phase of the CuO planes becomes unstable at high temperatures, when entropy effects lead to stripe meandering and eventually destroy the 1D stripes [12], and at high doping levels, when the expected distance between the stripes becomes so small that the Mott insulator phase between stripes collapses, thus recovering the 2D regime [2]. As a result, in optimally doped and overdoped regime the superconducting transition takes place when the $CuO_2$ layer is in the regime of the 2D doped Heisenberg system. Therefore, the rather "fragile" 1D SL phase seems to exist only in underdoped cuprates in the temperature window between $T^*$ and $T_c$.

In conclusion, we have described the transport properties of the even-chain SL compound $Sr_{2.5}Ca_{11.5}Cu_{24}O_{41}$ under pressure in terms of the 1D quantum transport model. Taking into account a remarkable scaling behaviour of the resistivity of $YBa_2Cu_4O_8$ at $T < T^*$ and $Sr_{2.5}Ca_{11.5}Cu_{24}O_{41}$ we have assumed that the former is also in the 1D regime at temperatures $T_c < T < T^*$ and therefore *the pseudo-gap in underdoped high-$T_c$ cuprates seems to be the spin-gap in the even-chain SL's associated with the stripe formation in the CuO planes at $T < T^*$*.

This work was supported by the GOA and FWO-Vlaanderen programs. L.T. is a research fellow supported by the Flemish Institute for the Stimulation of the Scientific and Technological Research in Industry (IWT); J.V. is a Research Fellow of the Flemish FWO. The authors are thankful to Y. Bruynseraede for useful discussions.